\documentclass[11pt,twoside]{article}
\usepackage{acta-info}
\usepackage{euler}
\usepackage[latin2]{inputenc}
\usepackage[ruled, vlined, linesnumbered, procnumbered]{algorithm2e}
\usepackage{ltablex}

\newcommand{\vol}{\textbf{3,} 2 (2011) 192--204}   %% to be completed by the editor, do not modify it
\newcommand{\amss}{\amssubj{
05C85
}}
\newcommand{\CRs}{\CRsubj{
G.2.2
}}
\newcommand{\keyw}{\keywords{
debt clearing, dynamic graph
}}

\begin{document}

\title{%
The debts' clearing problem: a new approach
}

\maketitle

\oneauthor{

\href{http://cs.ubbcluj.ro/~patcas/}{Csaba P\u ATCA\c S}
}{
\href{http://cs.ubbcluj.ro/}{Babeş-Bolyai University \\ Cluj-Napoca}
}{
 \href{mailto:patcas@cs.ubbcluj.ro}{patcas@cs.ubbcluj.ro}
}

\short{%
C. P\u atca\c s
}{
The debts' clearing problem: a new approach
}

\begin{abstract}
The debts' clearing problem is about clearing all the debts in a
group of $n$ entities (e.g. persons, companies) using a minimal
number of money transaction operations. In our previous works we
studied the problem, gave a dynamic programming solution solving it and proved that it is NP-hard. In this paper we adapt the problem to dynamic graphs and give a data structure to solve it. Based on this data structure we develop a new algorithm, that improves our previous one for the static version of the problem.
\end{abstract}

\section{Introduction}

In \cite{debt} we studied the debts' clearing problem, and gave a dynamic programming solution using $\Theta(2^n)$ memory and running in time proportional to $3^n$. The problem statement is the following:

\emph{Let us consider a number of $n$ entities (e.g. persons,
companies), and a list of $m$ borrowings among these entities. A
borrowing can be described by three parameters: the index of the
borrower entity, the index of the lender entity and the amount of
money that was lent. The task is to find a minimal list of money
transactions that clears the debts formed among these $n$ entities
as a result of the $m$ borrowings made.}

\begin{example}
\label{ex1}
\begin{tabular}{|c|c|c|}
\hline
Borrower&Lender&Amount of money\\
\hline
1&2&10\\
2&3&5\\
3&1&5\\
1&4&5\\
4&5&10\\
\hline
\end{tabular}

\smallskip
Solution: \ \ 
\begin{tabular}{|c|c|c|}
\hline
Sender&Reciever&Amount of money\\
\hline
1&5&10\\
4&2&5\\
\hline
\end{tabular}

\end{example}

In \cite{debt} we modeled this problem using graph theory:

\begin{definition}
Let $G(V,A,W)$ be a directed, weighted multigraph without loops,
$|V|=n$, $|A|=m$, $W:A \rightarrow \mathbb{Z}$, where $V$ is the set
of vertices, $A$ is the set of arcs and $W$ is the weight function.
$G$ represents the borrowings made, so we will call it the
\textbf{borrowing graph}.
\end{definition}

\begin{figure}[h]
\begin{center}
\includegraphics[width=8cm]{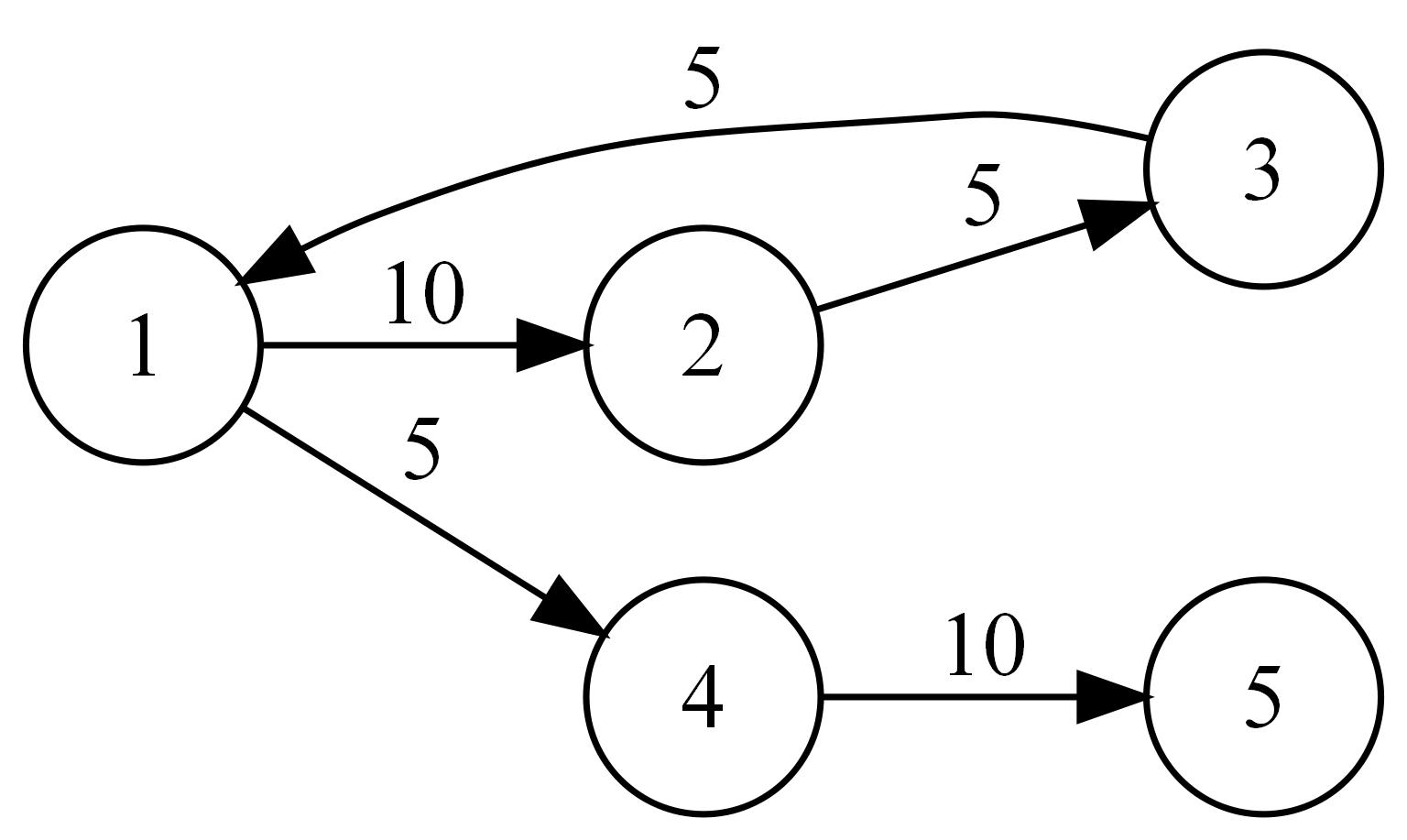}
\end{center}
\caption{The borrowing graph associated with the given example. An
arc from node $i$ to node $j$ with weight $w$ means, that entity $i$
must pay $w$ amount of money to entity $j$.}
\label{fig:borrowing}
\end{figure}

\begin{example}
The borrowing graph corresponding to Example \ref{ex1} is shown in Figure
\ref{fig:borrowing}.
\end{example}

\begin{definition}
Let us define for each vertex $v \in V$ the \textbf{absolute amount
of debt} over the graph $G$: 
$D_G(v) = \sum\limits_{\tiny\begin{array}{c}v' \in V\\ (v,v') \in A \end{array}}
W(v,v') - \sum \limits_{\tiny\begin{array}{c}v'' \in V\\ (v'',v) \in
A\end{array}} W(v'',v)$
\end{definition}

\begin{definition}
Let $G'(V,A',W')$ be a directed, weighted multigraph without loops,
with each arc $(i, j)$ representing a transaction of $W'(i, j)$
amount of money from entity $i$ to entity $j$. We will call this
graph a \textbf{transaction graph}. These transactions clear the
debts formed by the borrowings modeled by graph $G(V,A,W)$ if and
only if:

$D_G(v_i)=D_{G'}(v_i), \forall i=\overline{1,n}$, where
$V=\{v_1,v_2, \ldots, v_n\}$.

We will note this by: $G \sim G'$.
\end{definition}

\begin{figure}[h]
\begin{center}
\includegraphics[width=10cm]{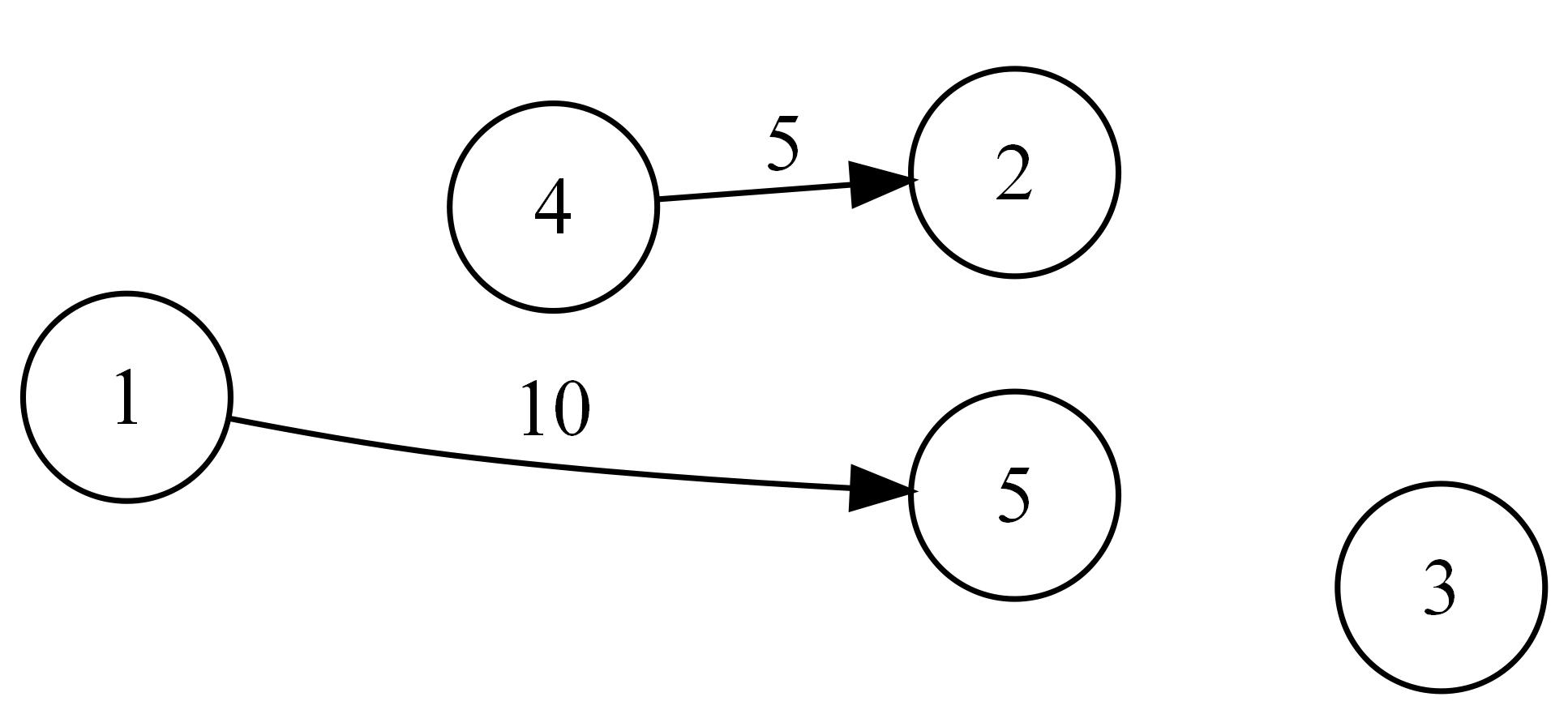}
\end{center}
\caption{The respective minimum transaction graph. An arc from node
$i$ to node $j$ with weight $w$ means, that entity $i$ pays $w$
amount of money to entity $j$.}
\label{fig:mintrans}
\end{figure}

\begin{example}
See Figure \ref{fig:mintrans} for a transaction graph with minimal
number of arcs corres\-ponding to Example \ref{ex1}.
\end{example}

Using the terms defined above, the debt's clearing problem can be
reformulated as follows:

\emph{Given a borrowing graph $G(V,A,W)$ we are looking for a
minimal tran\-sac\-tion graph $G_{min}(V,A_{min},W_{min})$, so that
$G \sim G_{min}$ and $\forall G'(V,A',W') :$ $G\sim G',
|A_{min}|\leq|A'|$ holds.}

\section{The debts' clearing problem in dynamic graphs}

\begin{definition}
A \textbf{dynamic graph} is a graph, that changes in time, by
undergoing a sequence of updates. An update is an operation, that
inserts or deletes edges or nodes of the graph, or changes
attributes associated to edges or nodes.
\end{definition}

In a typical dynamic graph problem one would like to answer queries
regarding the state of the graph in the current time moment. A good
dynamic graph algorithm will update the solution efficiently,
instead of recomputing it from scratch after each update, using the
corresponding static algorithm \cite{DemFinIta04}.

In the dynamic debts' clearing problem we want to support the following operations:

\begin{itemize}
\item \textsc{InsertNode$(u)$} -- adds a new node $u$ to the borrowing graph.
\item \textsc{RemoveNode$(u)$} -- removes node $u$ from the borrowing graph. In order for a node to be removed, all of its debts must be cleared first. In order to affect the other nodes as little as possible, the debts of $u$ will be cleared in a way that affects the least number of nodes, without compromising the optimal solution for the whole graph.
\item \textsc{InsertArc$(u, v, x)$} -- insert an arc in the borrowing graph. That is, $u$ must pay $x$ amount of money to $v$.
\item \textsc{RemoveArc$(u, v)$} -- removes the debt between $u$ and $v$.
\item \textsc{Query()} -- returns a minimal transaction graph.
\end{itemize}

\begin{figure}[h]
\begin{center}
\includegraphics[width=8cm]{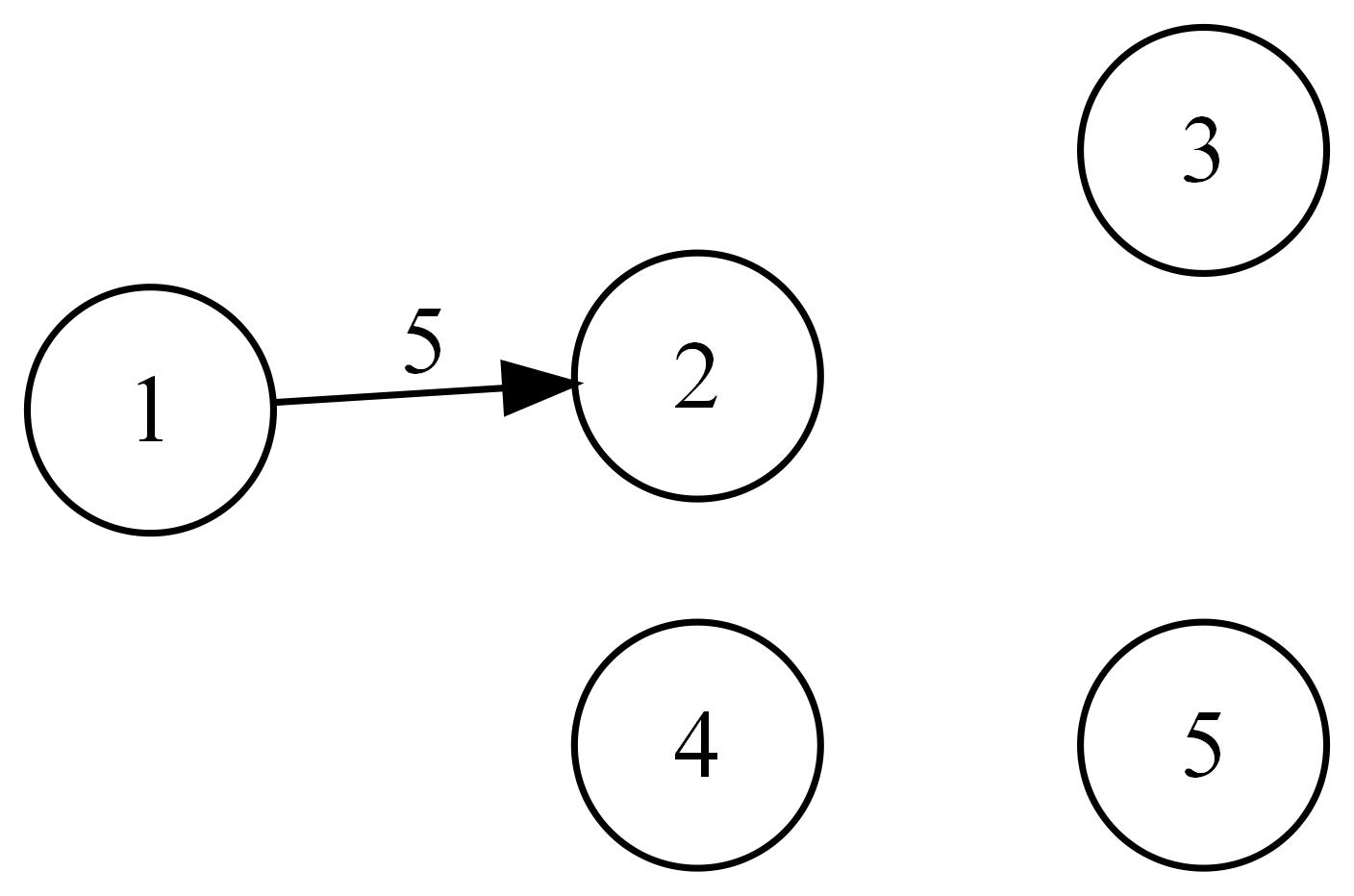}
\end{center}
\caption{Result of the \textsc{Query} operation called after the third arc was added}
\label{fig:query}
\end{figure}

\begin{example}
For instance calling the \textsc{Query} operation after adding the third arc in the borrowing graph corresponding to Example \ref{ex1} would result in the minimal transaction graph from Figure \ref{fig:query}.
\end{example}

These operations could be useful in the implementation of an application that facilitates borrowing operations among entities, such as BillMonk \cite{billmonk} or Expensure \cite{expensure}. When a new user registers to the system, it is equivalent with an \mbox{\textsc{InsertNode}} operation, and when a user wants to leave the system it is the same as a \mbox{\textsc{RemoveNode}}. When a borrowing is made, it can be implemented by a simple call of \textsc{InsertArc}. Two persons may decide, that they no longer owe each other anything. In this case \textsc{RemoveArc} can be useful. If the whole group decides, that it is time to settle all the debts, the \textsc{Query} operation will be used.

\section{A data structure for solving dynamic debts' clearing}

\label{sect:dynamic}

As the static version of the problem is NP-hard \cite{debt2}, it is not possible to support all these operations in polynomial time (unless P = NP). Otherwise we could just build up the whole graph one arc at a time, by $m$ calls of \textsc{InsertArc}, then construct a minimal transaction graph by a call of \textsc{Query}, which would lead to a polynomial algorithm for the static problem.

Our data structure used to support these operations is based on maintaining the subset of nodes, that have non-zero absolute amount of debt $V^* = \{u | D(u) \neq 0\}$. The sum of $D$ values for all the $2^{|V^*|}$ subsets of $V^*$ is also stored in a hash table called $sums$.

\subsection{InsertNode}

As for our data structure only nodes having non-zero $D$ values are important, and a new node will always start with no debts, it means that nothing has to be done when calling \textsc{InsertNode}.

\subsection{InsertArc}

When \textsc{InsertArc} is called, the $D$ values of the two nodes change, so $V^*$ can also change. When a node leaves $V^*$, we do not care about the updating the sum of the subsets it is contained in, because when a new node enters $V^*$ we will have to calculate the sum of all of the subsets it is contained in anyway.

If both $u$ and $v$ were in $V^*$ and remained in it after changing the $D$ values, then we simply add $x$ to the sum of all subsets containing $u$, but not $v$, and subtract $x$ from those containing $v$ but not $u$. The sum of the subsets containing both nodes does not change.

If one of the nodes was just added to $V^*$ ($D[u] = x$, or $D[v] = -x$), then all the sums of the subsets containing it must be recalculated. This recalculation can be done in $O(1)$ for each subset, taking advantage of sums already calculated for smaller subsets.

\begin{procedure}[H]
\caption{UpdateSums($u, v, x$)}
\tcp{Updates the sum of all subsets containing $u$ but not $v$}
\ForEach{$S \subset V^*$, such that $u \in S$ and $v \not \in S$}
{
  \lIf{D[u] = x}
  {
    $sums[S] := sums[S \setminus \{u\}] + x$\;
  }
  \lElse
  {
    $sums[S] := sums[S] + x$\;
  }
}
\end{procedure}

\begin{algorithm}[H]
\caption{\textsc{InsertArc$(u, v, x)$}}
\lIf{$D[u] = 0$}
{
  $V^* := V^* \cup \{u\}$\;
}
\lIf{$D[v] = 0$}
{
  $V^* := V^* \cup \{v\}$\;
}
$D[u] := D[u] + x$; $D[v] := D[v] - x$\;
\lIf{$D[u] = 0$}
{
  $V^* := V^* \setminus \{u\}$\;
}
\lIf{$D[v] = 0$}
{
  $V^* := V^* \setminus \{v\}$\;
}
\lIf{$D[u] \neq 0$}
{
  \UpdateSums($u, v, x$)\;
}
\lIf{$D[v] \neq 0$}
{
  \UpdateSums($v, u, -x$)\;
}
\If{$D[u] = x$ or $D[v] = -x$}
{
  \ForEach{$S \subset V^*$, such that $u, v \in S$}
  {
    $sums[S] := sums[S \setminus \{u, v\}] + D[u] + D[v]$\;
  }
}
\end{algorithm}

One call of UpdateSums iterates over $2^{|V^*| - 2}$ subsets, thus lines 6 and 7 of \textsc{InsertArc} together take $2^{|V^*| - 1}$ steps. Additionally line 9 takes $2^{|V^*| - 2}$ more steps.

\subsection{Query}

To carry out \textsc{Query} we observe, that finding a minimal transaction graph is equivalent to partitioning $V^*$ in a maximal number of disjoint zero-sum subsets, more formally $V^* = P_1 \cup \ldots \cup P_{max}, sums[P_i] = 0, \forall i = \overline{1, max}$ and $P_i \cap P_j = \emptyset, \forall i, j = \overline{1, max}, i \neq j$. The reason for this is, that all the debts in a zero-sum subset $P_i$ can be cleared by $|P_i| - 1$ transactions (see \cite{debt, debt2, verhoeff}), thus to clear all the debts, $|V^*| - max$ transactions are necessary. 
Let $S^0$ be the set of all subsets of $V^*$, having zero sum: $S^0 = \{S | S  \subset V^*,  sums[S] = 0\}$. Then, to find the maximal partition, we will use dynamic programming.

Let $dp[S]$ be the maximal number of zero-sum sets, $S \subset V^*$ can be partitioned in.

$dp[S] = \left\{
\begin{tabular}{cl}
not defined, & if $sums[S] \neq 0$\\
0, & if $S = \emptyset$\\
$\max\{dp[S \setminus S'] + 1 | S' \subset S, S' \in S^0\}$, & otherwise.\\
\end{tabular}
\right.$

\medskip
Building $dp$ takes at most $2 ^ {|V^*|} \cdot |S^0|$ steps.

As the speed at which \textsc{Query} can be carried out depends greatly on the size of $S^0$, we can use two heuristics to reduces its size, without compromising the optimal solution. To facilitate the running time of these heuristics, $S^0$ can be implemented as a linked list.

\paragraph{Clear pairs} Choosing sets containing exactly two elements in the partition will never lead to a suboptimal solution, if the remaining elements are partitioned correctly \cite{verhoeff}. Thus, before building $dp$, sets having two elements can be removed from $S^0$, along with all the sets, that contain those two elements (because we already added them to the solution, so there is no need to consider sets that contain them in the dynamic programming): $S^0 := S^0 \setminus (\{u, v\} \cup \{S' | u \in S'$ or $v \in S'\})$.

\begin{procedure}
\caption{ClearPairs()}
$max := 0$\;
$inPair := \emptyset$\;
\ForEach{$S \in S^0$}
{
  \If{$|S| = 2$}
  {
    \If{$S \cap inPair = \emptyset$}
    {
      $max := max + 1$\;
      $P_{max} := S$\;
    }
    $inPair := inPair \cup S$ \;
  }
}
\ForEach{$S \in S^0$}
{
  \lIf{$|S| \cap inPair \neq \emptyset$}
  {
    $S^0 := S^0 \setminus S$\;
  }
}
\end{procedure}

The running time of this heuristic is $\Theta(|S^0|)$.

\paragraph{Clear non-atomic sets} If a set $S_i \in S^0$ is contained in another set $S_j \in S^0$, then $S_j$ can be safely discarded, because $S_j \setminus S_i$ will also be part of $S^0$, and combining $S_i$ with $S_j \setminus S_i$ always leads to a better solution, than using $S_j$ alone: $S^0 := S^0 \setminus \{S_j | \exists S_i \in S^0: S_i \subset S_j\}$.

\begin{procedure}
\caption{ClearNonAtomic()}
\ForEach{$S_i \in S^0$}
{
  \ForEach{$S_j \in S^0, S_i \neq S_j$}
  {
    \lIf{$S_i \subset S_j$}
    {
      $S^0 := S^0 \setminus S_j$\;
    }
  }
}
\end{procedure}

This heuristic can be carried out in $\Theta(|S^0|^2)$.

\subsection{RemoveNode}

To delete a node $u$ with the conditions listed in the introduction is equivalent to finding a set $P$ of minimal cardinality containing $u$, that can still be part of an optimal partition, that is $dp[V^*] = dp[V^* \setminus P] + 1$. This algorithm can not be used together with the \textbf{Clear pairs} heuristic, because clearing pairs may compromise the optimal removal of $u$. The running time is the same as for \textsc{Query}, because $dp$ must be built.

\subsection{RemoveArc}

Because clearing an arc between two nodes is the same as adding an arc in the opposite direction, this can be easily implemented using \textsc{InsertArc}. If the $D$ values of the two nodes have the same sign, it means, that no arc could appear in a minimal transaction between the two nodes, so nothing has to be done.

\begin{algorithm}[H]
\caption{\textsc{RemoveArc$(u, v)$}}
\lIf{$D[u] < 0$ and $D[v] > 0$}
{
  \textsc{InsertArc$(u, v, \min\{-D[u], D[v]\})$}\;
}
\Else
{
  \lIf{$D[u] > 0$ and $D[v] < 0$}
  {
    \textsc{InsertArc$(v, u, \min\{D[u], -D[v]\})$}\;
  }
}
\end{algorithm}

\subsection{Implementation details}

In our implementation we used 32-bit integers to represent subsets. A subset of at most 32 nodes can be codified by a 32-bit integer by looking at its binary representation: node $i$ is in the subset if and only if the $i^{th}$ bit is one. This idea allows using bit operations to improve the running time of the program.

Because we did not use test cases having more than 20 nodes, the hash table $sums$ was implemented as a simple array having $2^n$ elements. $V^*$ was stored as an ordered array, but other representations are also possible, because the running time of the operations on $V^*$ is dominated by other calculations in our algorithm.

Before using the methods of the data structure for the first time, the memory for its data fields containing $V^*$ and $sums$ should be allocated and their values initialized, both being empty at the beginning. In a destructor type method these memory fields can be deallocated.

\section{A new algorithm for the static problem}

\label{sect:newstatic}

We can observe, that the \textsc{Query} operation needs only the set $S^0$ to be built, and in order to build $S^0$ the sum of all subsets of $V^*$ needs to be calculated. Thus, after processing all the arcs in $\Theta(m)$ time and finding the $D$ values, we build $V^*$ in $\Theta(n)$ time along with the $sums$ hash table, that can be built in $\Theta(2^{|V^*|})$ by dynamic programming:

$sums[S = \{s_1, \ldots s_k\}] = \left\{
\begin{tabular}{cl}
0, & if $S = \emptyset$\\
$D[s_1]$, & if $|S| = 1$\\
$sums[\{s_2, \ldots s_k\}] + D[s_1]$, & otherwise.\\
\end{tabular}
\right.$

After $sums$ is built, we can construct $S^0$ by simply iterating once again over all the subsets of $V^*$ and adding zero-sum subsets to $S^0$. Then we clear pairs and non-atomic sets, call \textsc{Query} and we are done. This yields to a total complexity of $\Theta(m + n + 2^{|V^*|} + |S^0|^2 + 2 ^ {|V^*|} \cdot |S^0|)$.

\section{Practical behavior}

As it can be seen from the time complexities of the operations, the behavior of the presented algorithms depends on the cardinalities of $V^*$ and $S^0$ and their running times may vary from case to case.

We have made some experiments to compare our new algorithms and the static algorithm presented in \cite{debt}. We used the same 15 test cases which were used, when the problem was proposed in 2008 at the qualification contest of the Romanian national team. Figure \ref{fig:tests} contains the structure of the graphs used for each test case.

\begin{figure}[t]
\begin{tabularx}{12cm}{|c|c|c|c|p{8cm}|}
\hline
Test & $n$ & $m$ & $|A_{min}|$ & Short description\\
\hline
1 & 20 & 19 & 1 & A path with the same weight on each arc\\
2 & 20 & 20 & 0 & A cycle with the same weight on each arc\\
3 & 8 & 7 & 7 & Minimal transaction graph equals to borrowing graph\\
4 & 20 & 19 & 19 & Two connected stars\\
5 & 20 & 15 & 15 & Yields to $D[i] = 2, \forall i = \overline{1, 10}$, $D[i] = -1, \forall i = \overline{11, 19}$ and  $D[20] = -11$, maximizing the number of triples (zero-sets with cardinality three)\\
6 & 20 & 10 & 10 & Yields to $D[i] = 99, \forall i = \overline{1, 10}$, $D[i] = -99, \forall i = \overline{11, 20}$, maximizing the number of pairs\\
7 & 20 & 19 & 12 & A path with random weights having close values ($50 \pm 10$)\\
8 & 20 & 20 & 10 & A cycle with random weights having close values ($50 \pm 10$)\\
9 & 10 & 100 & 7 & Random graph with weights $\leq 10$\\
10 & 12 & 100 & 9 & Random graph with weights $\leq 10$\\
11 & 15 & 100 & 11 & Random graph with weights $\leq 10$\\
12 & 20 & 100 & 14 & Random graph with weights $\leq 10$\\
13 & 20 & 19 & 15 & A path with consecutive weights\\
14 & 20 & 30 & 15 & Ten pairs, a path, a star and triples put together\\
15 & 20 & 100 & 15 & Dense graph with weights $\leq 3$\\
\hline
\end{tabularx}
\caption{The structure of the test cases}
\label{fig:tests}
\end{figure}

In our first experiment we compared three algorithms: the old static algorithm based on dynamic programming from \cite{debt}, our new static algorithm described in Section \ref{sect:newstatic} and the dynamic graph algorithm based on the data structure presented in Section \ref{sect:dynamic}. For the third algorithm we called \textsc{InsertArc} for each arc, then \textsc{Query} once in the end, after all arcs were added.

\begin{figure}[t]
\begin{tabularx}{16cm}{|c|c|c|c|c|}
\hline
Test & Old static & New static & Dynamic & Improvement\\
& algorithm & algorithm & algorithm & \\
\hline
1 & 0.018 & \textbf{0.017} & 0.018 & 3.7\%\\
2 & 0.019 & \textbf{0.007} & 0.010 & 64.4\%\\
3 & 0.013 & \textbf{0.007} & 0.007 & 43.9\%\\
4 & \textbf{0.036} & 0.050 & 0.441 & -38.1\%\\
5 & 6.383 & \textbf{0.551} & 0.932 & 91.3\%\\
6 & \textbf{0.013} & 0.138 & 0.488 & -909.7\%\\
7 & \textbf{0.014} & 0.034 & 0.169 & -145.2\%\\
8 & \textbf{0.015} & 0.019 & 0.084 & -26.6\%\\
9 & 0.014 & \textbf{0.008} & 0.010 & 45.5\%\\
10 & 0.014 & \textbf{0.007} & 0.022 & 47.6\%\\
11 & 0.015 & \textbf{0.008} & 0.106 & 44.4\%\\
12 & \textbf{0.016} & 0.056 & 5.526 & -242.8\%\\
13 & 0.765 & \textbf{0.079} & 0.465 & 89.6\%\\
14 & \textbf{0.013} & 0.048 & 1.527 & -271.7\%\\
15 & 2.274 & \textbf{0.218} & 8.553 & 90.3\%\\
\hline
\end{tabularx}
\caption{Average running times for the first experiment, all given in seconds. The best running times are bolded for each test. The last column shows the improvement of the new static algorithm over the old one in percentage. A negative value means, that no improvement was done.}
\label{fig:exp1}
\end{figure}

\begin{figure}[t]
\begin{tabularx}{16cm}{|c|c|c|c|c|c|c|}
\hline
Test & Old  & Dynamic & Improvement & $\overline{|V^*|}$ & $\overline{|S^0|}$ & $\overline{|S^0|}$ \\
& static & algorithm & & & & after\\
& algorithm & & & & & heuristics \\
\hline
1 & 0.079 & \textbf{0.019} & 76.0\%   & 2 & 1 & 0\\
2 & 0.066 & \textbf{0.013} & 79.8\%   & 1.9 & 0.95 & 0\\
3 & 0.029 & \textbf{0.009} & 69.6\%   & 5 & 1 & 0.85\\
4 & \textbf{0.109} & 0.588 & -437.8\% & 11 & 1 & 0.94\\
5 & 10.541 & \textbf{1.515} & 85.6\%  & 11.66 & 7257 & 407.26\\
6 & \textbf{0.040} & 0.469 & -1072.5\%& 11 & 25094.2 & 0\\
7 & \textbf{0.063} & 0.217 & -243.6\% & 10.15 & 1035.63 & 3.57\\
8 & \textbf{0.066} & 0.142 & -113.5\% & 10.75 & 801.3 & 7.7\\
9 & 0.294 & \textbf{0.017} & 94.1\%   & 9.35 & 10.4 & 4.2\\
10 & 0.300 & \textbf{0.046} & 84.4\%  & 11.28 & 40.28 & 13.17\\
11 & 0.329 & \textbf{0.258} & 21.6\%  & 13.59 & 283.9 & 33.19\\
12 & \textbf{1.575} & 11.346 & -620.0\% & 17.72 & 6002.9 & 189.38 \\
13 & 1.101 & \textbf{0.588} & 46.5\%  & 11 & 969.947 & 80.94\\
14 & \textbf{0.105} & 2.801 & -2551.4\%& 16.46 & 1428.6 & 6.36 \\
15 & \textbf{340.719} & 611.282 & -79.4\%& 18.26 & 11790.8 & 9501.37\\
\hline
\end{tabularx}
\caption{Average running times for the second experiment, all given in seconds. The best running times are bolded for each test. The third column shows the improvement of the dynamic algorithm over recomputing from scratch with the old static one in percentage. A negative value means, that no improvement was done. The last three columns show the average cardinality of $V^*$, $S^0$ and $S^0$ after applying both heuristics respectively.}
\label{fig:exp2}
\end{figure}

We executed each algorithm three times for each test case, and computed the average of the running times. The new static algorithm was the fastest in nine test cases, while the old static algorithm was the fastest in the remaining six test cases. Looking at the average running time over all the test cases, the new static algorithm was clearly the fastest with an average of 0.08 seconds. The old static algorithm came second with 0.64 seconds, and the dynamic algorithm third with 1.22 seconds. The difference between the last two is surprisingly small, taking into account that the dynamic algorithm may perform $2^n$ steps after each arc insertion. Running times are shown in Figure \ref{fig:exp1}.

In the second experiment we used the same methodology to compare our new dynamic algorithm and the static algorithm presented in \cite{debt}. For the first algorithm the solution was recomputed from scratch each time an arc was read from the input file, and for the second after each \textsc{InsertArc} a \textsc{Query} was also executed. The dynamic algorithm was faster for eight test cases, recalculating from scratch was faster in the other seven cases. The average running time over all test cases is 23.6 seconds for the first algorithm and 41.9 seconds for the dynamic algorithm, mostly due to the last test case which runs for a long time compared to the others. Without taking into account the last test case the average running times are 1.05 and 1.28 seconds respectively.

By comparing the last two columns of the table depicted in Figure \ref{fig:exp2}, one can see how powerful our heuristics are, reducing the cardinality of $S^0$ by several magnitudes in many cases. We can observe, that the dynamic algorithm usually performs slower than recomputing from scratch, when the size of $S^0$ before applying the heuristics is quite large, at least several hundreds. The reason behind this is probably the quadratic complexity of clearing non-atomic sets.

\section{Conclusions}

In this paper we introduced a new data structure capable of supporting arc insertions and deletions, node insertions and deletions in a dynamic borrowing graph, along with finding the minimal transaction graph. Using this data structure we developed a new static algorithm, which is faster than the one described in \cite{debt} in many cases and in average.

We find the running times of the dynamic algorithm and recomputing from scratch with the old static algorithm to be comparable on average. With a good heuristic, that runs in reasonable time, but still reduces the size of $S^0$ significantly, a better performance could be possible for the dynamic algorithm. Finding such a heuristic remains an open problem.

Our experiments are not meant to be an exact comparison among the algorithms, as the running time can greatly depend on the details of the implementation. Their purpose was just to get a general overview on the behavior of the various algorithms for different kind of graphs.

\bigskip
\rightline{\emph{Received: May 16, 2011 {\tiny \raisebox{2pt}{$\bullet$\!}} Revised: October 10, 2011}} %% to be completed by the editor

\end{document}